\documentclass{elsart}

\usepackage{graphicx}

\usepackage{amsmath}
\usepackage{amssymb}

\begin{document}

\begin{frontmatter}



\title{New cellular automaton designed to simulate epitaxial films growth}


\author{Rafa{\l} Kosturek}
\and
\author{Krzysztof Malarz\corauthref{km}}
\ead{malarz@agh.edu.pl}
\corauth[km]{Corresponding author.
Fax: +48 12 6340010.} 

\address{Department of Applied Computer Science,
Faculty of Physics and Nuclear Techniques,
AGH University of Science and Technology\\
al. Mickiewicza 30, PL-30059 Krak\'ow, Poland}

\begin{abstract}
In this paper a simple (2+1) solid-on-solid model of the epitaxial films growth based on random deposition followed by breaking particle-particle lateral bonds and particles surface diffusion is introduced.
The influence of the critical number of the particle-particle lateral bonds $z$ and the deposition rate on the surface roughness dynamics and possible surface morphology anisotropy is presented.
The roughness exponent $\alpha$ and the growth exponent $\beta$ are $(0.863,0.357)$, $(0.215,0.123)$, $(0.101,0.0405)$ and $(0.0718,0.0228)$ for $z=1$, 2, 3 and 4, respectively.

Snapshots from simulations of the growth process are included.
\end{abstract}

\end{frontmatter}


\section{Introduction}
Theoretical modelling of films epitaxial growth --- which is growth of an oriented single-crystal film of one material upon a single-crystal substrate of another \cite{MBE} and when the main microscopic process is particles deposition followed by their diffusion on the surface --- may be grouped to continuum and discrete approaches \cite{fractal}.

The continuum ones are based on the stochastic differential Langevin equations \cite{ew,kpz,lai,park} which group models to the various classes of universality associated with the some set of critical exponents.
In discrete approach computational methods may be applied, with {\em cellular automata technique} (CA) \cite{CA,ilachinski} among others.
The latter bases usually on the Arrhenius-like kinetics \cite{arrhenius,ferrando} and requires knowing many parameters, for instance:
\begin{itemize}
\item material and neighbourhood-dependent activation energies for different elementary processes,
\item adatom/adatom, substrate-atom/substrate-atom and substrate-atom/adatom bonds strengths,
\item vibration factors,
\item substrate temperature,
\item incoming particles flux, etc.
\end{itemize}
Usually, the values of these parameters are fitted to reproduce some experimentally measured particles and/or surface characteristics.

The title of Ref. \cite{ilachinski} characterise the CA technique very well, as for CA, both, discrete time and space are necessary.
The model must include also the rule which tells how the states of lattice cells are subsequently updated.
Usually, CA models for epitaxial growth simulations \cite{sos} base on simple or even toy, mechanical rules, e.g. random deposition followed by particles relaxation (see Ref. \cite{fractal} for review).
In the relaxation process particles often virtually move to the nearest-neighbourhoods (NN) sites to check offered there accommodation conditions and then chose the best one.
Quite often, the subsequent particle arrives to the place of its first contact with surface only when previously deposited particle migration process has been completed.
Such a situation corresponds to very low flux of incoming  particles and does not meet real condition during molecular beam epitaxy experiments.

Here we would like to present simple CA which looses both: disadvantages of non-physical particles virtual movement to the NN sites and low particles flux but still does not need material dependent constants.

\section{Model}
Presented here model, as usually, is extension to random deposition model (RDM) with additional particles relaxation.
The solid-on-solid approximation is applied, so the film may be fully characterised by single-valued function $h(x,y,t)$ of the film height in planar coordinates $(1\le x\le L, 1\le y\le L)$ in time $t$.

We start our simulation with perfectly flat substrate.
Every $\tau L^2$ time steps new jet of $\theta_{\text{dep}}L^2$ particles arrives.
Each time step --- between subsequent acts of the depositions --- particles `sitting' on the column top may diffuse on the surface.
The only mobile particles are those which currently have less than $z_x$ and $z_y$ created particle-particle lateral bonds (PPLB) in $x$- and $y$-direction, respectively.
For isotropic case only one number $z$ guards the particles mobility.
Active particles and their movement directions are picked up randomly.
The particles are not allowed to climb on higher levels, but they are able to jump down at the terrace edge.
The simulation is carried out until a desired film thickness $\theta_{\text{max}}$ has been deposited.

\section{Results}
We characterise surface morphology with some statistical parameters such as film thickness $\theta$ and surface width $w$.
The film roughness $w$ is defined as the film height $h$ standard deviation 
\[
w^2(t)=\sum_{x,y} [h(x,y,t)-\theta(t)]^2/L^2
\]
from the average film thickness 
\[
\theta(t)=\sum_{x,y} h(x,y,t)/L^2.
\]

For anisotropic case ($z_x\ne z_y$ and $z_x z_y\ne 0$) few measures $\varepsilon$ describing quantitatively surface morphology anisotropy were proposed and investigated.

\begin{figure}
\centering
(a) \includegraphics[width=.4\textwidth]{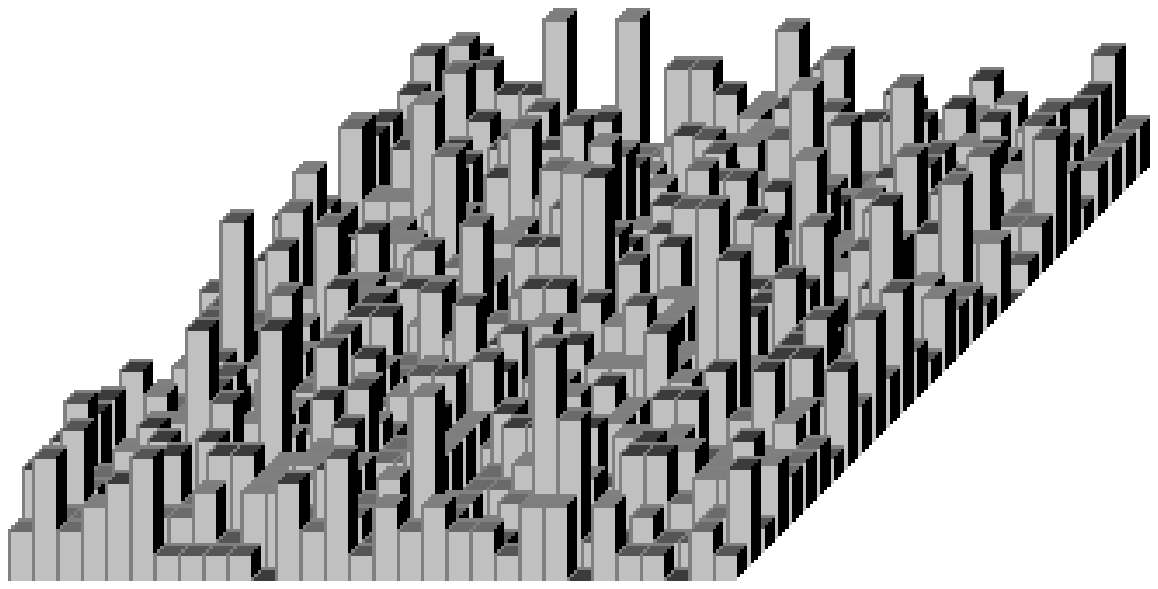}
(b) \includegraphics[width=.4\textwidth]{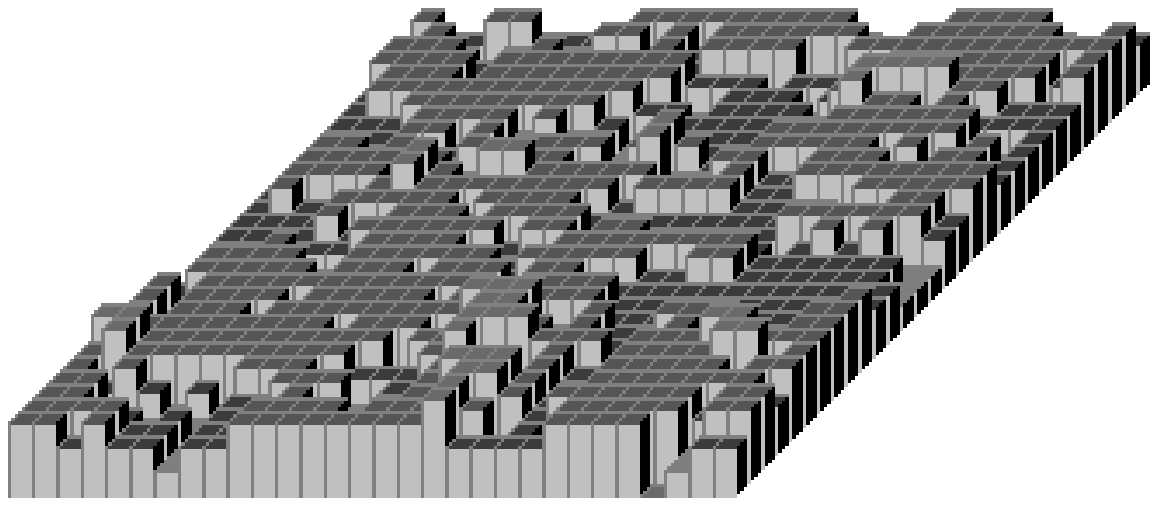}\\ 
(c) \includegraphics[width=.4\textwidth]{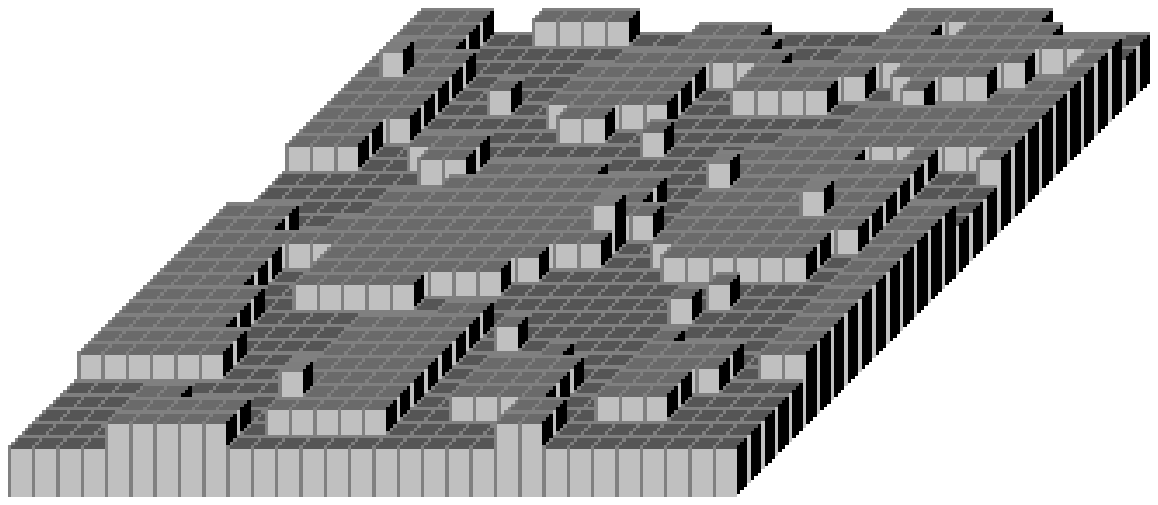}
(d) \includegraphics[width=.4\textwidth]{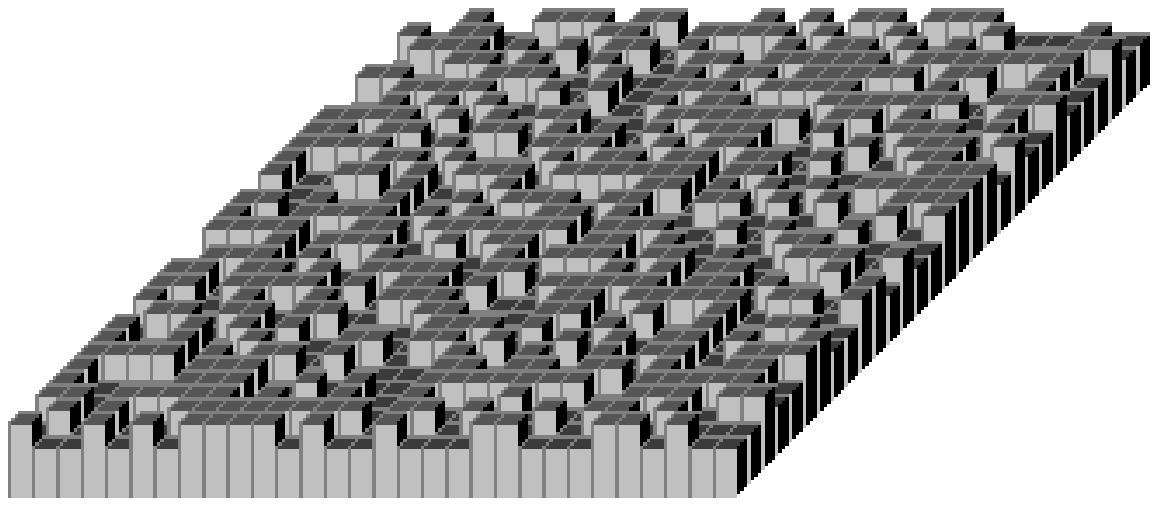}\\
(e) \includegraphics[width=.4\textwidth]{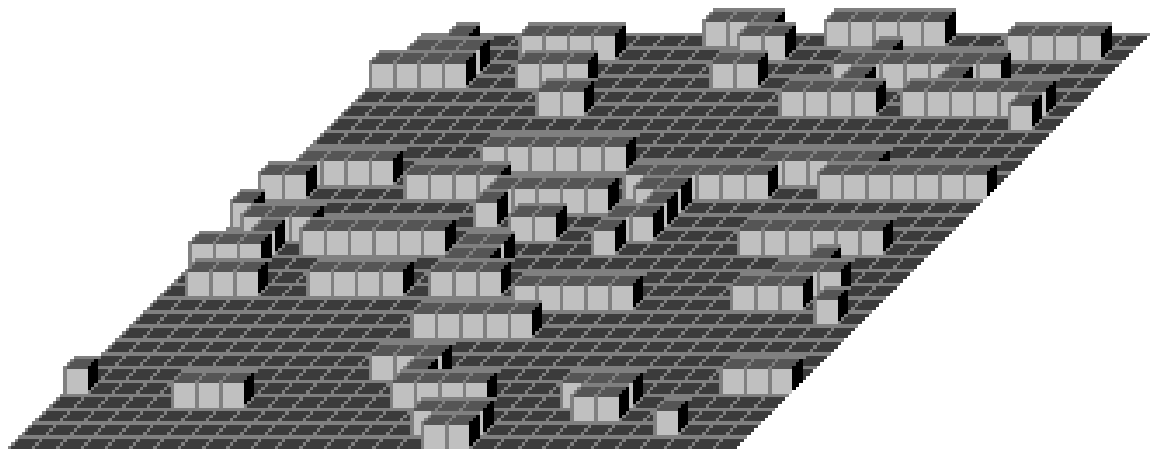}
(f) \includegraphics[width=.4\textwidth]{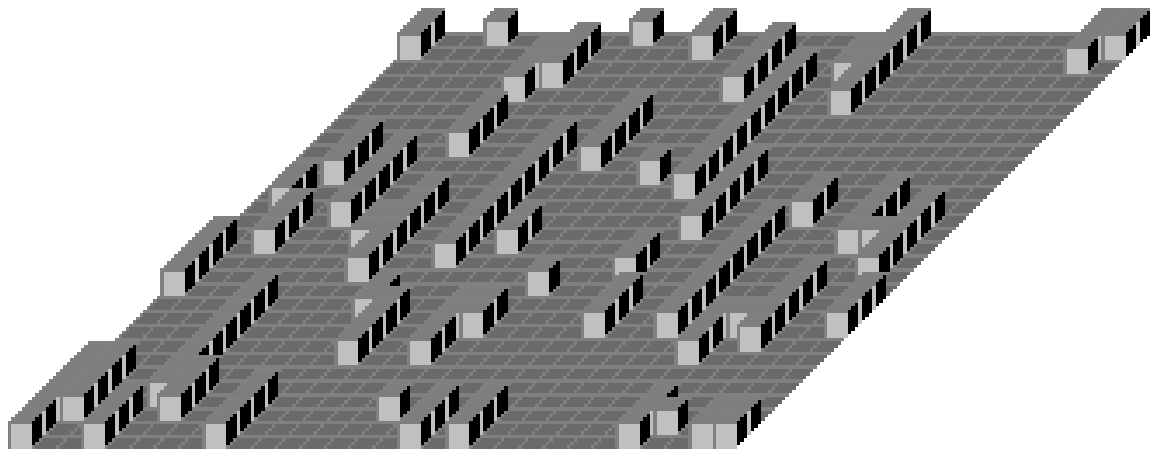}\\ 
(g) \includegraphics[width=.4\textwidth]{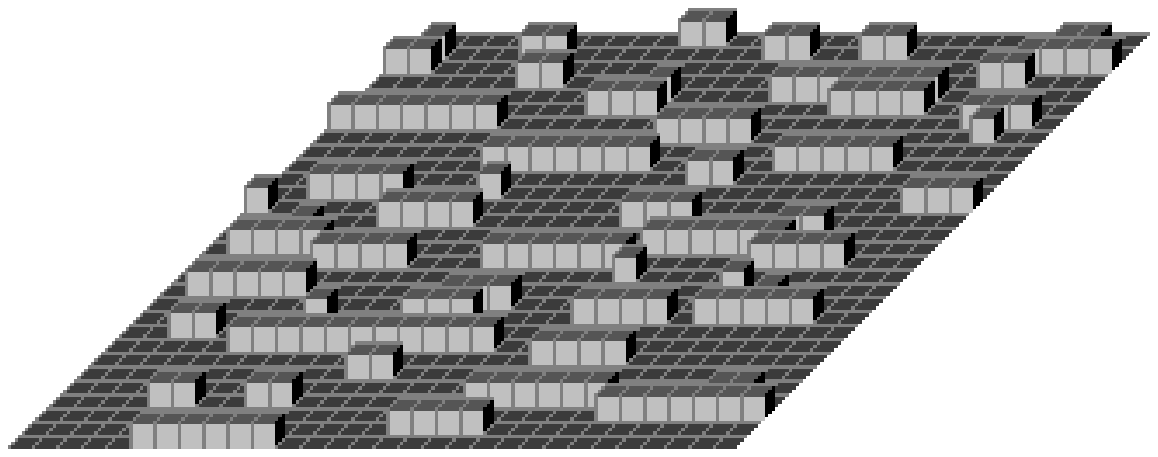}
(h) \includegraphics[width=.4\textwidth]{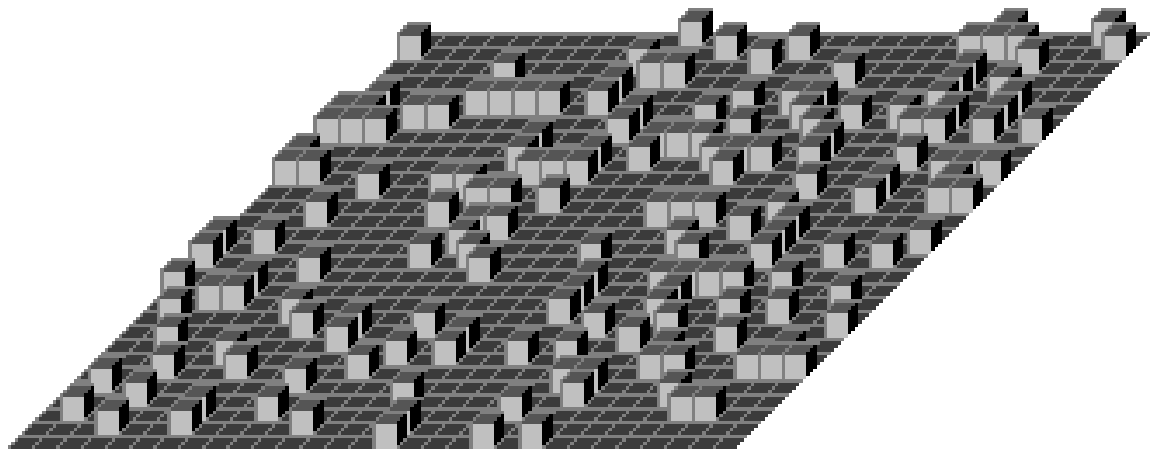}
\caption{
Samples of film for (a) RDM ($z z_x z_y=0$ and/or $\tau=0$), isotropic case with (b) $z=1$ (c) $z=2$ (d) $z=4$, and anisotropic cases (e) $z_x=1$, $z_y=2$ (f) $z_x=2$, $z_y=1$ (g) $z_x=1$, $z_y=3$ (h) $z_x=2$, $z_y=3$.
Average film height $\theta_{\text{max}}=2.5$ [ML] for RDM and isotropic cases and $\theta_{\text{max}}=0.2$ [ML] for anisotropic ones.
For all cases the deposition rate was given by $\theta_{\text{dep}}=0.01$ [ML] and $\tau=10$.
$30\times 30$-large parts of lattices are presented.
}
\label{fig-sample}
\end{figure}

For $z z_x z_y=0$ and/or $\tau=0$ the RDM results are well reproduced with Poisson distribution of film heights and $w(t)=\sqrt{\theta(t)}$.
The film sample for RDM is presented in Fig. \ref{fig-sample}(a).

\subsection{Isotropic case}
\begin{figure}
\centering
\includegraphics[width=.49\textwidth]{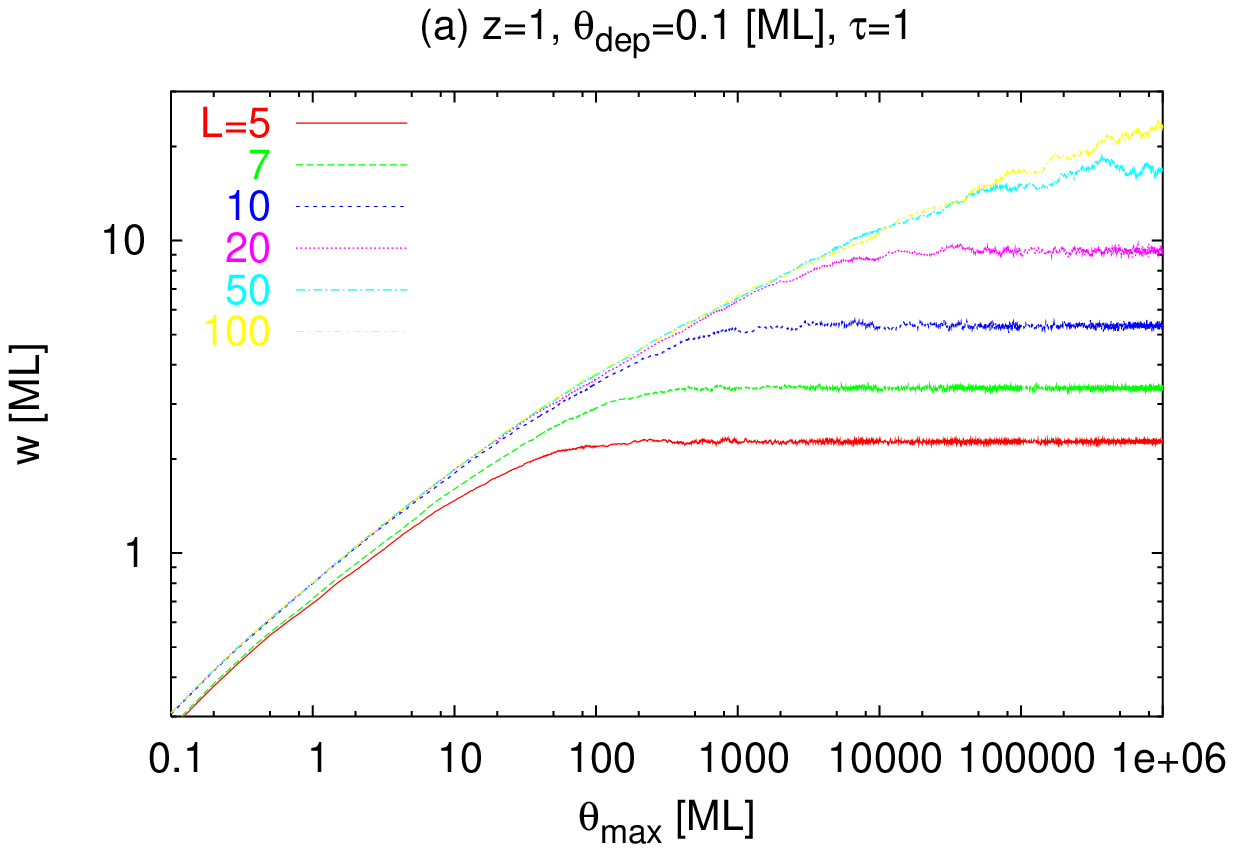}
\includegraphics[width=.49\textwidth]{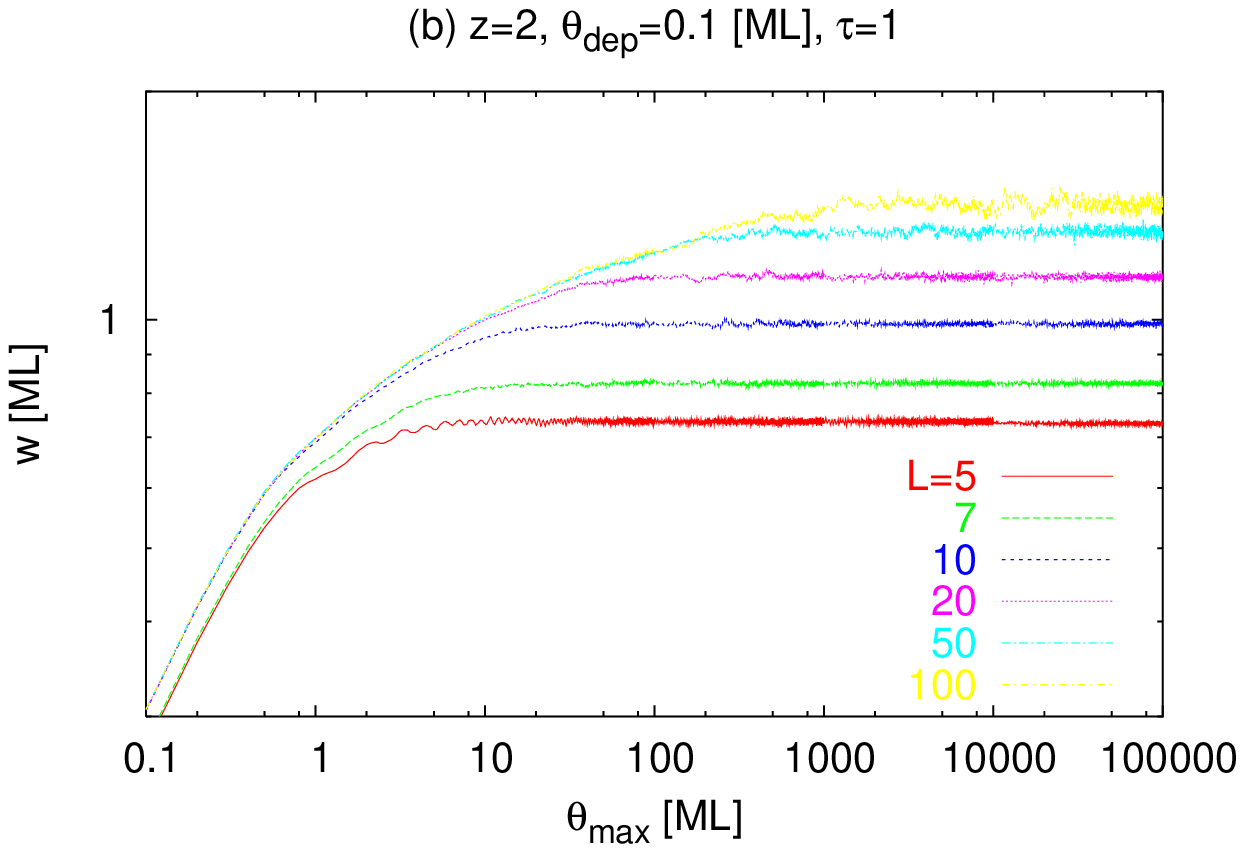}
\includegraphics[width=.49\textwidth]{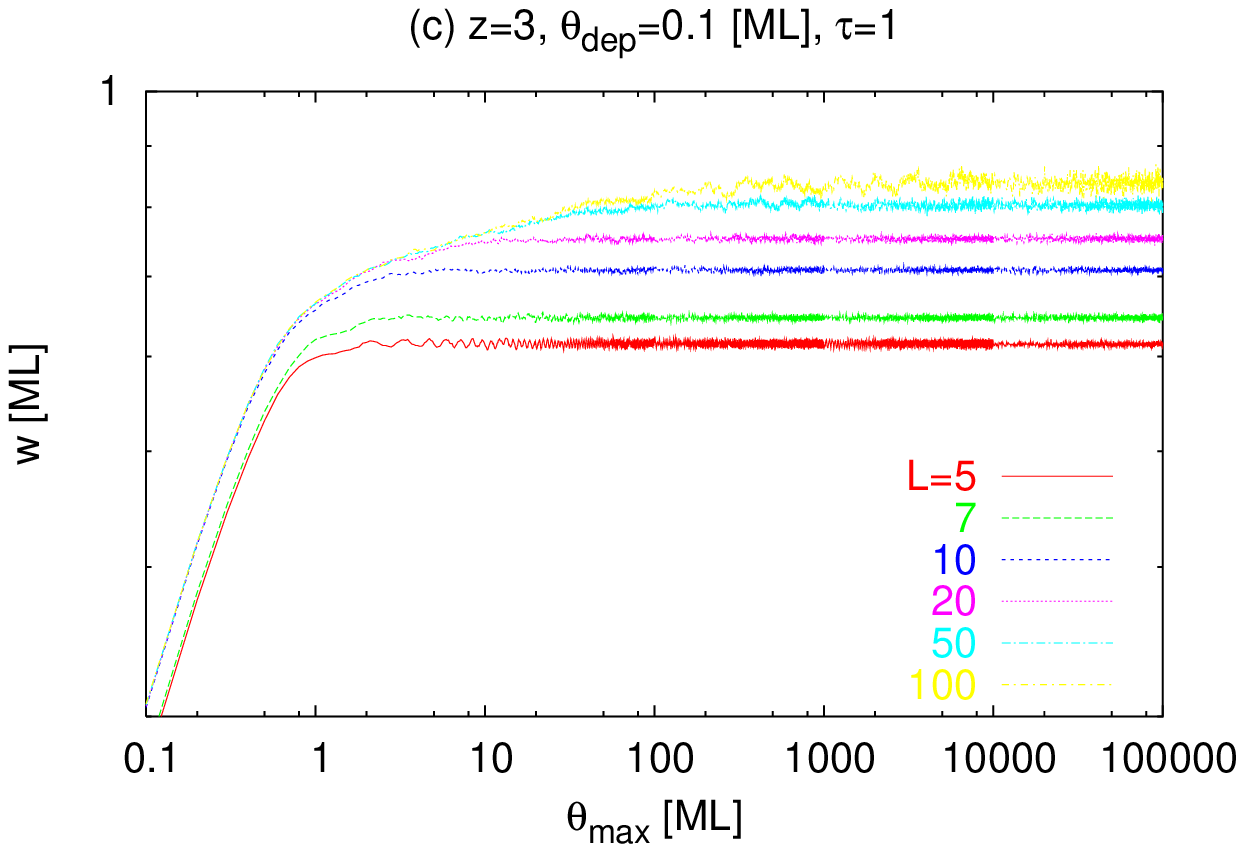}
\includegraphics[width=.49\textwidth]{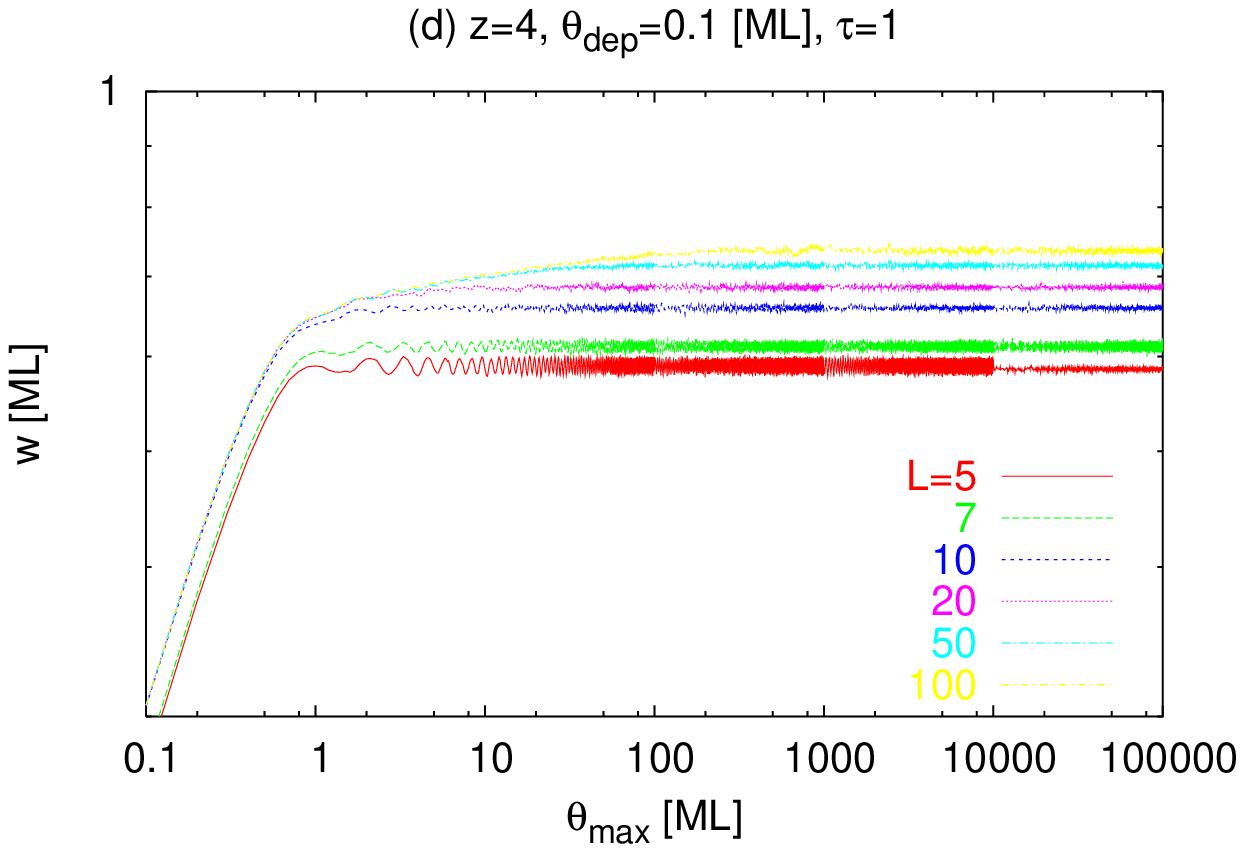}
\includegraphics[width=.49\textwidth]{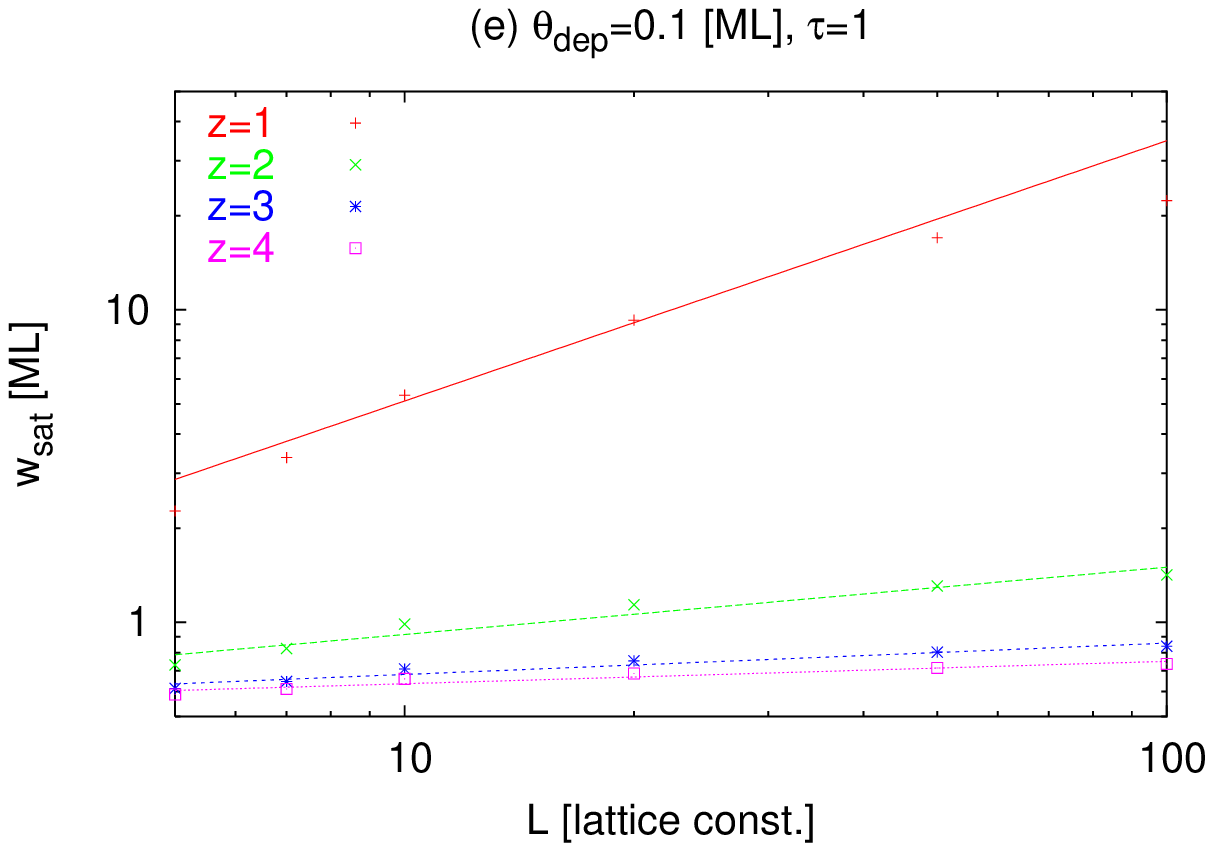}
\caption{Dependence of the roughness dynamics $w(t)$ for
(a) $z=1$,
(b) $z=2$,
(c) $z=3$,
(d) $z=4$
and
(e) the saturation level $w_{\text{sat}}$ on linear lattice size $L$
for $\theta_{\text{dep}}=0.1$ [ML] and $\tau=1$.}
\label{fig-fv}
\end{figure}

For wide variety of discrete models of surface growth the surface width $w$ is expected to follow the dynamical Family--Vicsek \cite{fv} scaling law:
\begin{subequations}
\label{eq-fv}
\begin{equation}
w(t,L) \propto L^\alpha\cdot f(t/L^\zeta),
\label{eq-fv-a}
\end{equation}
with scaling function
\begin{equation}
f(x)=
\begin{cases}
x^\beta\text{ and }\beta=\alpha/\zeta 	& \text{ for }x\ll 1,\\
1 					& \text{ for }x\gg 1,\\
\end{cases}
\label{eq-fv-b}
\end{equation}
\end{subequations}
where $\alpha$, $\beta$ and $\zeta$ are roughness, growth and dynamic exponent, respectively.
From Eq. \eqref{eq-fv} one may read surface roughness dynamics: before reaching the characteristic film thickness $\theta_{\text{sat}}$ the surface roughness grows like $w(t)\propto t^\beta$ (see Fig. \ref{fig-fv}(a) and Tab. \ref{tab-w}).
Then, the roughness saturates on the level $w_{\text{sat}}$ depending on the substrate linear size $L$: $w_{\text{sat}}\propto L^\alpha$ (see Fig. \ref{fig-fv}(b) and Tab. \ref{tab-w}).
These exponent predict also classes of universality \cite{park,ts}.
However, determination of the exponents $\alpha$, $\beta$ and $\zeta$ solely on the calculation of $w(L,t)$ has to be treated with caution \cite{siegert}.

\begin{figure}
\centering
\includegraphics[width=.49\textwidth]{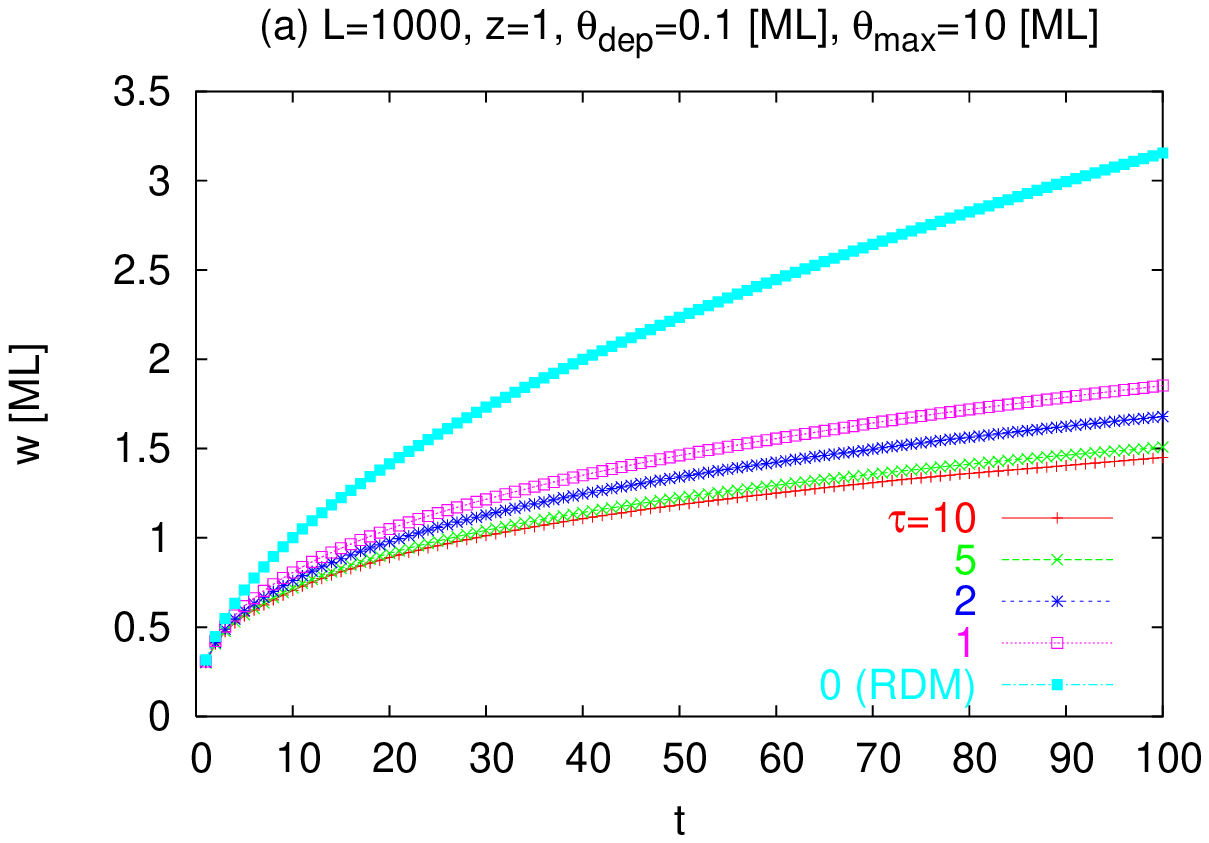}
\includegraphics[width=.49\textwidth]{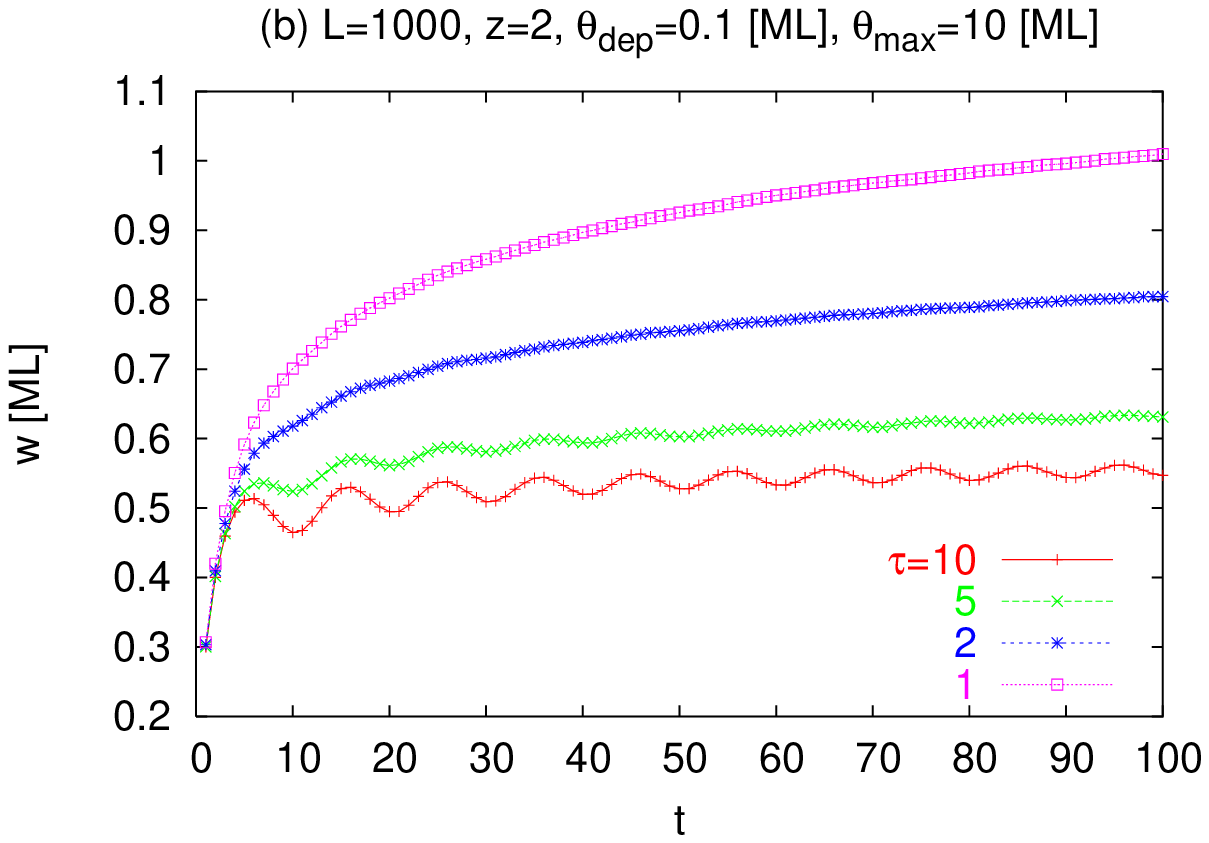}
\includegraphics[width=.49\textwidth]{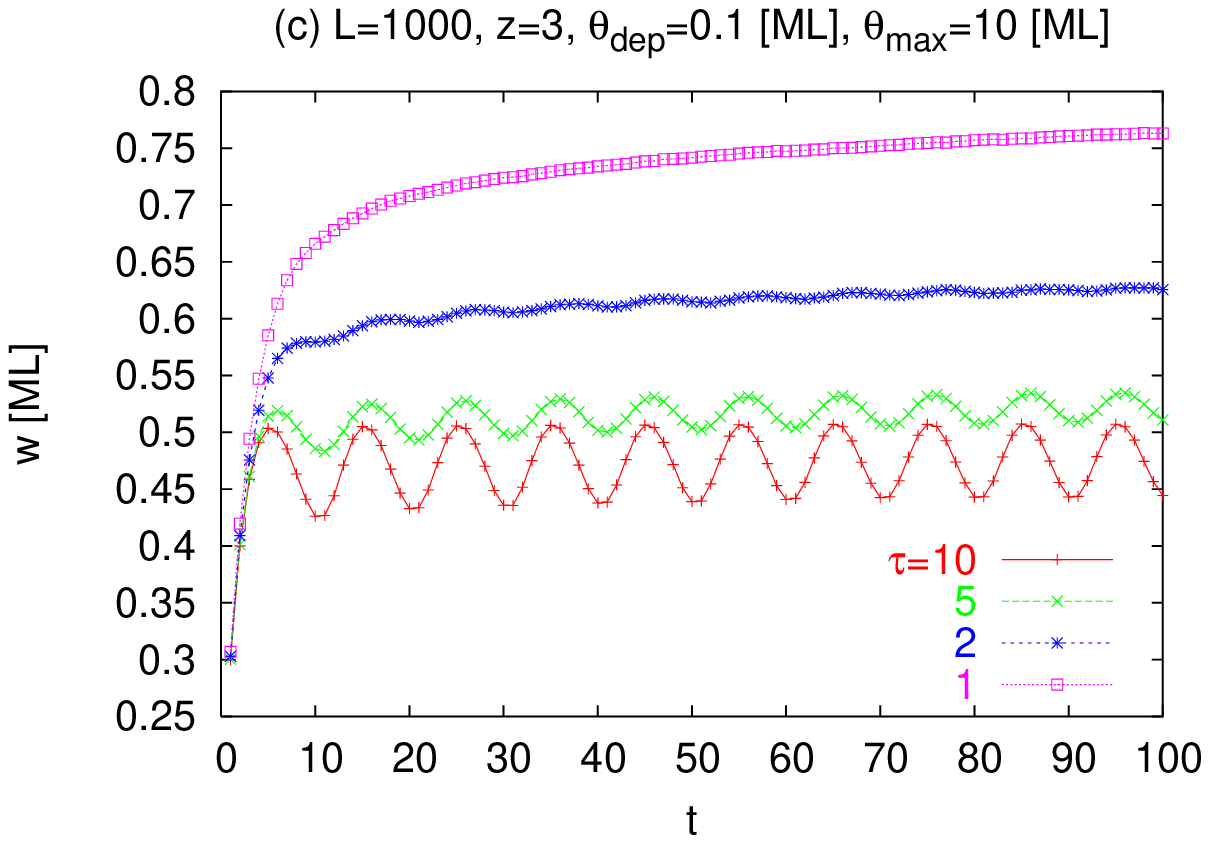}
\includegraphics[width=.49\textwidth]{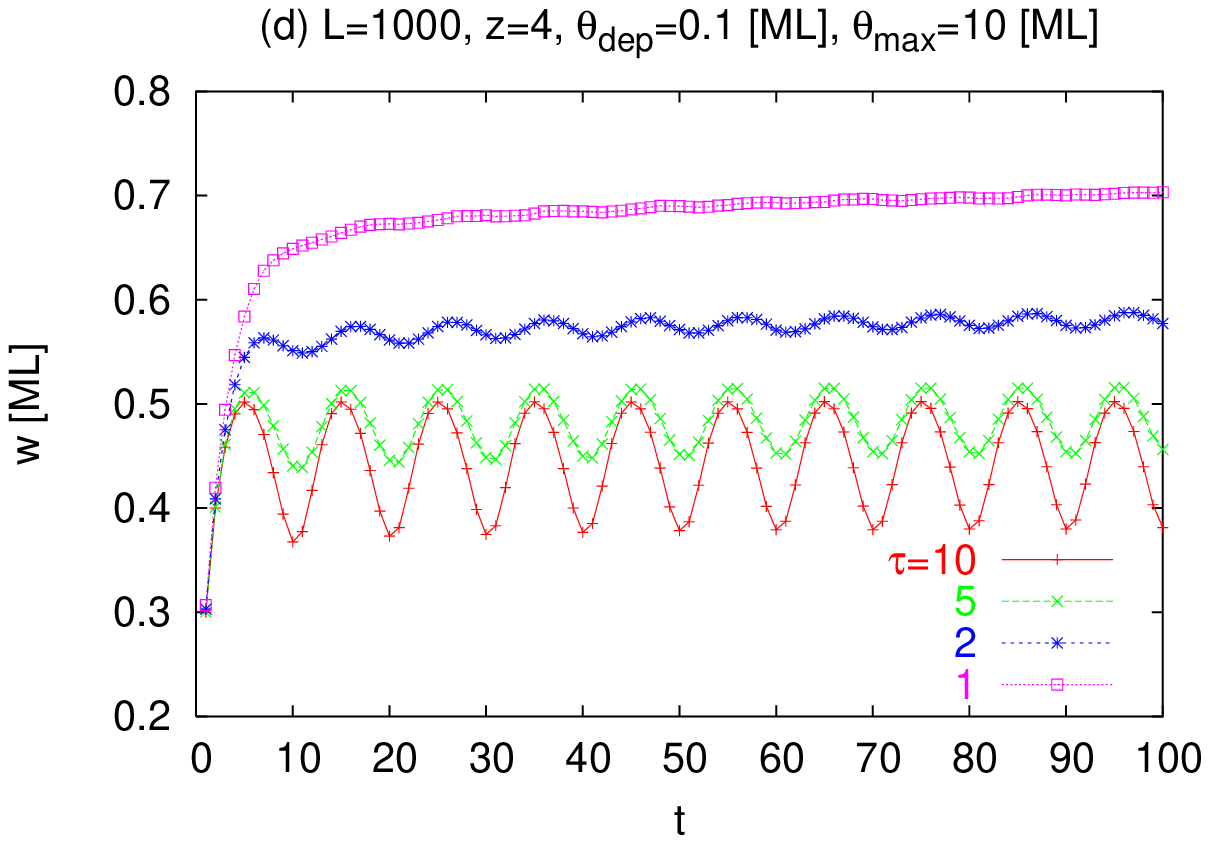}
\caption{Time evolution of the surface roughness $w(t)$ for different number of the particle relaxations $\tau$ and different critical values of PPLB
(a) $z=1$,
(b) $z=2$,
(c) $z=3$ and 
(d) $z=4$
($L=10^3$, $\theta_{\text{dep}}=0.1$ [ML]).}
\label{fig-w-z}
\end{figure}

\begin{figure}
\centering
\includegraphics[width=.49\textwidth]{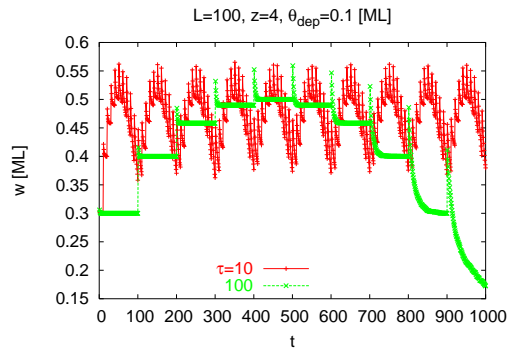}
\caption{Time evolution of the surface roughness $w(t)$ for $L=100$, $z=4$, $\theta_{\text{dep}}=0.1$ [ML] and different values of $\tau$.}
\label{fig-lupa}
\end{figure}

For large enough number of relaxations, e.g. $\tau\ge 5$, the characteristic layer-by-layer oscillations of roughness $w(t)$ occur and Frank--van de Merwe growth mode is observed (see Figs. \ref{fig-w-z} and \ref{fig-lupa}).
The influence of the critical number of PPLB $z$ on the roughness and growth exponents for given $\theta_{\text{dep}}=0.1$ [ML] and $\tau=1$ are collected in Tab. \ref{tab-w}.
\begin{table}
\centering
\caption{Roughness $\alpha$ and growth exponents $\beta$ dependence on the critical PPLB number $z$ for $\theta_{\text{dep}}=0.1$ [ML] and $\tau=1$.
The results are average over $N_{\text{run}}$ independent simulations (from several thousand for $L=5$ to a few for $L=100$).}
\label{tab-w}
\begin{tabular}{ccccc}
\hline
$z$      & 1     	& 2		& 3		& 4 \\
\hline
$\alpha$ & 0.863	& 0.215		& 0.1005	& 0.0718 \\
$\beta$  & 0.357 	& 0.123 	& 0.0405  	& 0.0228 \\
\hline
\end{tabular}
\end{table}
\begin{figure}
\centering
\includegraphics[width=.49\textwidth]{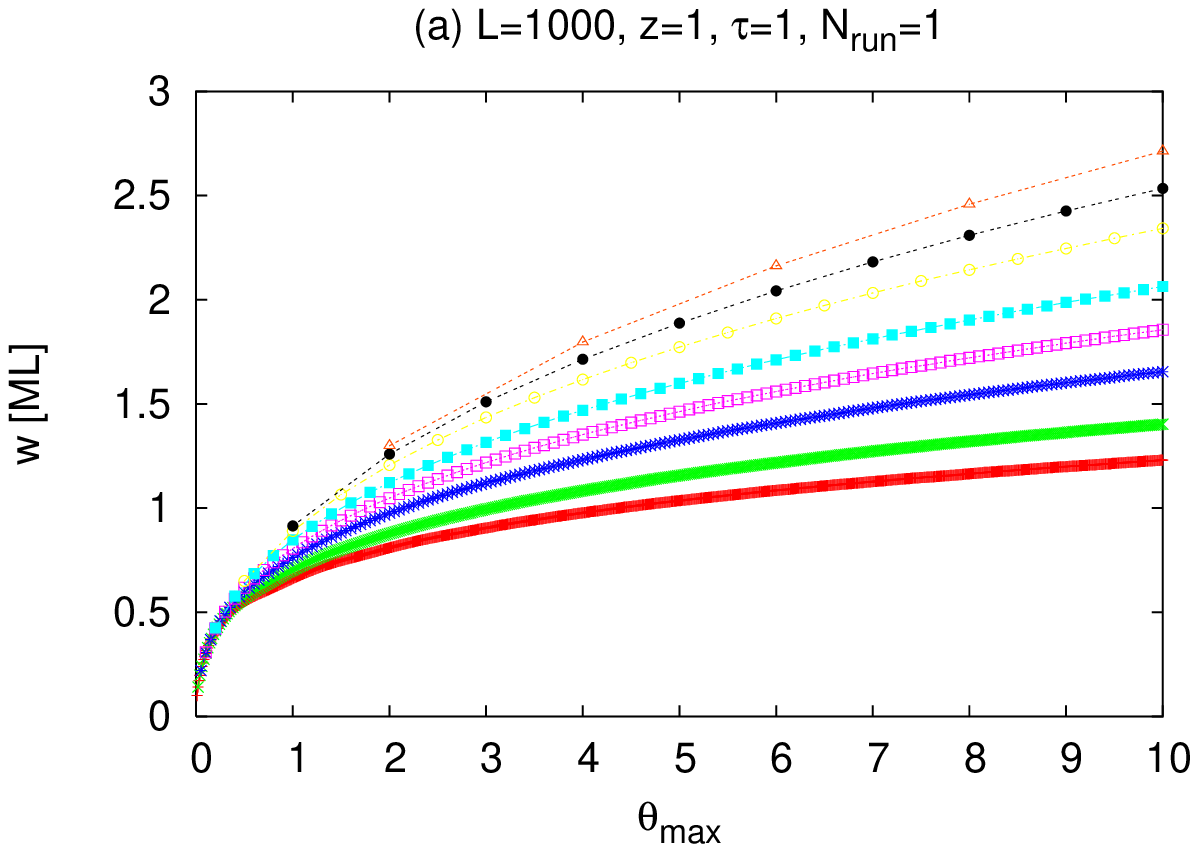}
\includegraphics[width=.49\textwidth]{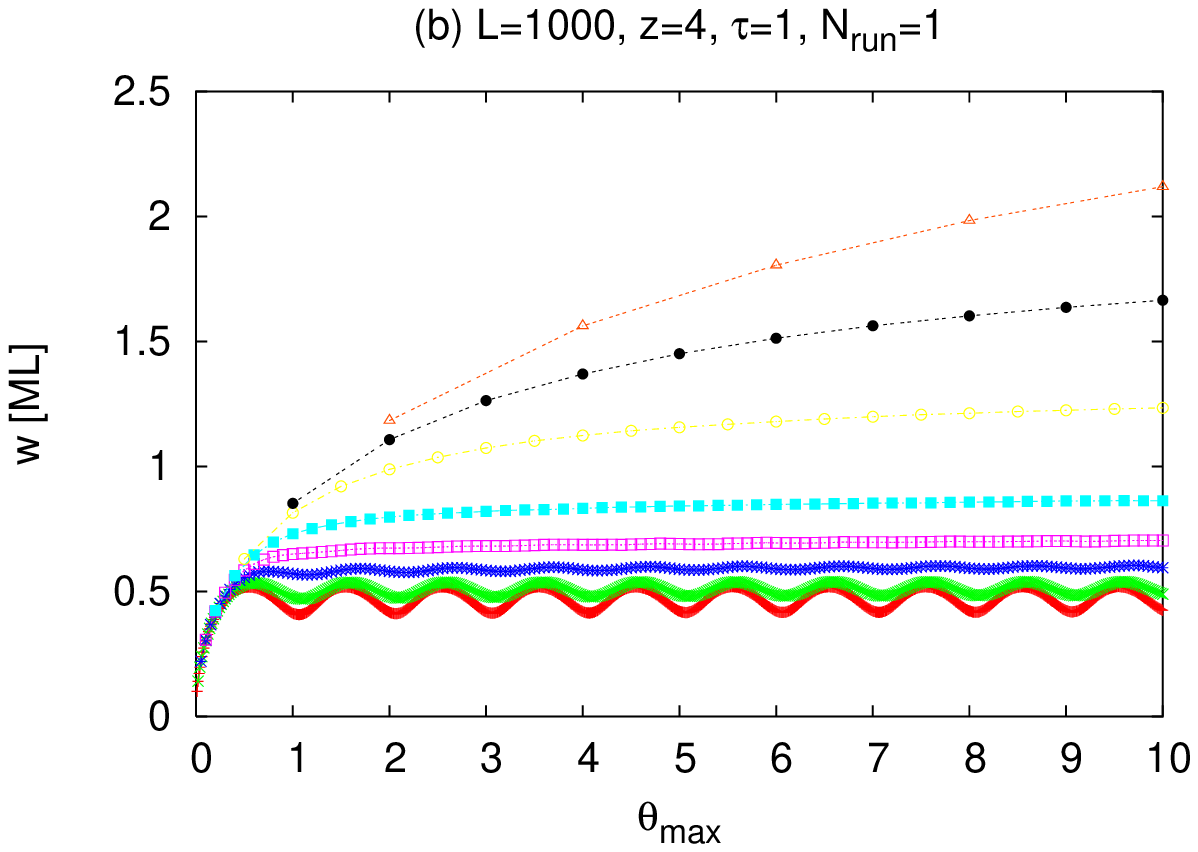}
\caption{Dependence of the roughness dynamics $w(t)$ on $\theta_{\text{dep}}$ for $L=1000$, $\tau=1$ and (a) $z=1$ (b) $z=4$.
$\theta_{\text{dep}}=0.01$, 0.02, 0.05, 0.1, 0.2, 0.5, 1.0, 2.0 from bottom to top.}
\label{fig-theta-dep}
\end{figure}

Surface roughness decreases with $\tau$ (see Fig. \ref{fig-w-z}) but increases with $\theta_{\text{dep}}$ (see Fig. \ref{fig-theta-dep}), as expected.

The film samples for $z=1, 2, 4$ are presented in Figs. \ref{fig-sample}(b-d).

\subsection{Anisotropic case}
\label{sec-aniso}
For submonolayer substrate coverage (e.g. $\theta_{\text{max}}=0.2$ [ML]), and when $z_x\ne z_y$ and $z_x z_y\ne 0$ (see snapshots from simulations presented in Figs. \ref{fig-sample}(e-h)) we search for quantitative measure of surface morphology anisotropy.
One of such characteristic --- based on the height-height correlation function --- was proposed in Ref. \cite{malarz99b} and employed for investigation of the surface morphology thermal evolution in Ref. \cite{malarz00}.

Here, for quantitative characterisation of the surface morphology anisotropy we propose set of $\varepsilon$ parameters:
$\varepsilon_1=(\phi_x-\phi_y)/(\phi_x+\phi_y)$, 
$\varepsilon_2=\phi_x/\phi_y$, 
and $\varepsilon_3=\ell/A$, where $\phi_x$ and $\phi_y$ are $x$- and $y$-side of the minimal rectangle which totally covers whole cluster, $\ell$ is the cluster perimeter and $A$ is the the cluster area.
The periodic boundary condition modification of the Hoshen--Kopelman algorithm \cite{hka} for cluster labelling were used.
The results are averaged over all existing clusters.
\begin{table}
\centering
\caption{Different measures of the surface morphology anisotropy $\varepsilon$ for $L=1000$, $\theta_{\text{dep}}=0.01$ [ML], $\theta_{\text{max}}=0.2$ [ML] and $\tau=10$.}
\label{tab-epsilon}
\begin{tabular}{ccccccc}
\hline
$z_x$ 		&1     & 2       &1     &3        &2      &3        \\
$z_y$ 		&2     & 1       &3     &1        &3      &2        \\
\hline
$\varepsilon_1$ &0.424 &$-0.424$ &0.457 &$-0.457$ &0.013 &$-0.013$\\
$\varepsilon_2$ &3.05  &0.443    &3.32  &0.408    &1.11   &1.04     \\
$\varepsilon_3$ &2.61  &2.61     &2.61  &2.61     &3.56   &3.56     \\
\hline
\end{tabular}
\end{table}
Note, that for isotropic case $\phi_x=\phi_y$ yields $\varepsilon_1=0$ and  $\varepsilon_2=1$.
The values of $\varepsilon_3$ depend on particular clusters shape and their size, i.e. $\varepsilon_3=4/a$ for square of side $a$, $\varepsilon_3=2/r$ for circle with radius $r$ and $\varepsilon_3=4\sqrt{3}/a$ for equilateral triangle of side $a$.
It seems that $\varepsilon_1$, for which $\varepsilon_1(z_x,z_y)=-\varepsilon_1(z_y,z_x)$, is much more suitable for quantitative morphology characterisation than $\varepsilon_2$ and/or $\varepsilon_3$.

\section{Summary}
For all values of the critical PPLB $z$ we observe scaling of the surface roughness $w(L,t)$ according to Family--Vicsek law \eqref{eq-fv} with characteristic exponents given in Tab. \ref{tab-w}.

For fixed set of model control parameters the roughness $w$ decreases both with parameter $\tau$ and $z$.
The latter one may be also seen in a very naive way as an equivalent of the absolute substrate temperature, as with increase of $z$ the particles become more and more mobile (they stick at the place of deposition for $z=0$, while for $z=4$ their moves become completely forbidden only when they have four NN).

The increase of the parameter $\tau$ corresponds to effective decrease of the incoming particles flux per unit substrate area and the time step.
For lower particles flux we observe smoother surfaces, as in between subsequent acts of deposition particles may make more moves.
The similar effect may be reproduced with direct decreasing of the deposition rate $\theta_{\text{dep}}$, and thus surface roughness $w$ increases with $\theta_{\text{dep}}$.

For anisotropic case the dependence of $\varepsilon_1$ parameter may be qualitatively compared with results of Monte Carlo simulations from Ref. \cite{ferrando}, CA approach presented in Refs. \cite{malarz99b,malarz00} and --- at last but not least --- with experimental data \cite{ferrando}.
So, when one identify $z$ with the substrate temperature one may observe initial increase of $\varepsilon_1$ followed by its decrease --- i.e. $\varepsilon_1(\bar z=1.5) < \varepsilon_1(\bar z=2) > \varepsilon_1(\bar z=2.5)$, where $\bar z=(z_x+z_y)/2$.
We conclude, that $\varepsilon_1$ has similar properties as $\varepsilon$ defined via height-height correlation function in Ref. \cite{malarz00}, and additionally is antisymmetric (i.e. $\varepsilon_1(z_x,z_y)=-\varepsilon_1(z_y,z_x)$) and vanishes for isotropic case ($z_x=z_y$ and/or $z_x z_y=0$).

Presented here asynchronous and probabilistic CA 
\begin{itemize}
\item does not need virtual movements of the particles to the NN sites to check if they offer better accommodation condition (i.e. higher coordination number or lower column height),
\item and do not have many activation energies (for different elementary processes like surface diffusion or breaking bonds) and lattice vibration factor, etc.
\end{itemize}
The latter may be seen as the model disadvantage, however, one can easily imagine how to introduce material constants into growth rules to obtain the model which lays somewhere in between CA with simple, mechanical rules of surface relaxation and full reversible Arrhenius based kinetic Monte Carlo model.

The program for growth process microscopy is available from our web-page \cite{www}.

\ack
The part of calculations was carried out in ACC\---CY\-FRO\-NET\---AGH.
The machine time on HP Integrity Superdome is financed by the Ministry of Science and Information Technology in Poland under grant No. MNiI/\-HP\_I\_SD/\-AGH/\-002/\-2004.



\end{document}